\documentclass[aps,floats,amssymb,twocolumn,superscriptaddress,preprintnumbers]{revtex4}

\usepackage{graphicx}
\usepackage{amssymb}

\newcommand{\fpi}{f_\pi}
\newcommand{\mpi}{m_\pi}
\newcommand{\gev}{\ {\rm GeV}}
\newcommand{\mev}{\ {\rm MeV}}
\newcommand{\fm}{\ {\rm fm}}
\newcommand{\non}{\nonumber}

\newcommand{\wk}{\omega_k}
\newcommand{\chpt}{$\chi$PT}
\newcommand{\npr}{$\eta '$}
\newcommand{\NM}{\mu_N}
\newcommand{\C}{{\cal C}}

\newcommand{\eq}[1]{Eq.~(\ref{#1})}
\newcommand{\eqs}[2]{Eqs.~(\ref{#1}) and (\ref{#2})}

\begin{document}

\preprint{
\vbox{
\hbox{ADP-04-15/T597}
\hbox{JLAB-THY-04-22}
}}

\title{Leading Quenching Effects in the Proton Magnetic Moment}

\author{R. D. Young}
\affiliation{    Special Research Centre for the
                 Subatomic Structure of Matter,
                 and Department of Physics and Mathematical Physics,
                 University of Adelaide, Adelaide SA 5005,
                 Australia}
\author{D. B. Leinweber}
\affiliation{    Special Research Centre for the
                 Subatomic Structure of Matter,
                 and Department of Physics and Mathematical Physics,
                 University of Adelaide, Adelaide SA 5005,
                 Australia}
\author{A. W. Thomas}
\affiliation{    Special Research Centre for the
                 Subatomic Structure of Matter,
                 and Department of Physics and Mathematical Physics,
                 University of Adelaide, Adelaide SA 5005,
                 Australia}
\affiliation{Jefferson Laboratory, 12000 Jefferson Ave.,
             Newport News, VA 23606 USA}

\begin{abstract}
  
We present the first investigation of the extrapolation of quenched
nucleon magnetic moments in quenched chiral effective field theory.
We utilize established techniques in finite-range regularisation and
compare with standard dimensional regularisation methods.
Finite-volume corrections to the relevant loop integrals are also
addressed.  Finally, the contributions of dynamical sea quarks to the
proton moment are estimated using a recently discovered
phenomenological link between quenched and physical QCD.

\end{abstract}

\maketitle


\section{Introduction}
Describing the quark content of nucleon structure in terms of QCD is a
fundamental aim of modern nuclear physics.  As an inherently
nonperturbative theory, the most rigorous approach to low-energy
phenomena in QCD is by numerical simulations in lattice gauge
theory.  Although restricted at present to the regime of quark masses
exceeding those realized in nature, recent advances in effective field
theory (EFT) have made it possible to accurately extract the physical
nucleon mass from QCD \cite{Leinweber:2003dg}.

These advances and breakthroughs in chiral effective field theory
($\chi$EFT) have their origin in the study of a range of hadron
properties in QCD, including nucleon magnetic moments and charge radii
\cite{Leinweber:1998ej,Hackett-Jones:2000qk,Hemmert:2002uh,Hackett-Jones:2000js},
the nucleon sigma commutator
\cite{Leinweber:2000sa,Leinweber:2003dg,Procura:2003ig}, moments of
structure functions
\cite{Detmold:2001jb,Detmold:2002nf,Hemmert:2003cb} and the
$\rho$-meson mass \cite{Leinweber:2001ac}.

With the the most detailed studies being on the extrapolation of the
nucleon mass, it has been shown that the use of finite-range
regularisation (FRR) enables the most systematically accurate
connection of effective field theory and lattice simulation results
\cite{Leinweber:2003dg,Young:2002ib,Leinweber:1999ig}. Mathematically
equivalent to dimensional regularisation (DR) to any finite order, FRR
chiral effective field theory provides a resummation of the chiral
expansion with vastly improved convergence properties. Central to FRR
is the presence of a finite energy scale which may be used to optimize
the convergence properties of the truncated expansion. The success of
FRR-EFT is highlighted by the observation that the higher-order terms
of the traditional dimensionally regulated expansion, although
individually large, must sum to zero to describe lattice QCD results.

Here FRR-$\chi$EFT is applied to nucleon magnetic moments calculated
in lattice QCD.  In particular, we investigate the modifications
required for the extrapolation of quenched simulation results.  Issues
with the formulation of EFT on a finite volume are also discussed.  We
consider the modifications to chiral loop integrals on a finite volume
and perform a fixed-volume extrapolation.

We extend a phenomenological link between quenched and dynamical
baryon masses \cite{Young:2002cj} to the case of magnetic moments to
estimate the artifacts associated with the quenched approximation.  We
find that the full QCD corrections of the quenched magnetic moments are
small over a wide range of quark mass.

Finally, the convergence properties of a truncated Taylor-series
expansion of the FRR results in powers of $\mpi$, analogous to that
encountered in dimensional regularisation, are investigated.  We
illustrate how any moderate truncation of the series expansion is
unable to connect with current lattice simulation results.  The
inclusion of higher-order chiral loop corrections associated with the
$\Delta$ baryon are also considered and found to be small.

%
\section{Magnetic Moments --- Quark Mass Dependence}

\begin{figure}[!t]
\begin{center}
\includegraphics[width=5cm, angle=0]{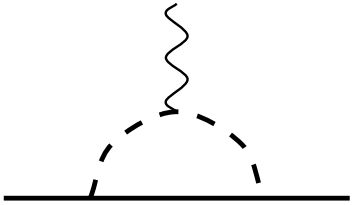}
\caption{Diagram providing the leading non-analytic contribution of
  pions (dashed curve) to the nucleon (solid line) magnetic moment.
\label{fig:piLNA}}
\end{center}
\end{figure}

In a general construction of an effective field theory for low-energy
QCD, the expansion of the nucleon's magnetic moment about the chiral
limit can be written as
\begin{eqnarray}
\mu_B &=& a^B_0 + a^B_2 \mpi^2 + a^B_4 \mpi^4 + \ldots \non\\
      & & \chi_{B\pi} I_\pi + \ldots \, .
\label{eq:exp}
\end{eqnarray}
The term $\chi_{B\pi} I_\pi$ denotes the leading nonanalytic (LNA)
chiral correction to the baryon magnetic moment of
Fig.~\ref{fig:piLNA}.  Coefficients of low-order nonanalytic
contributions to nucleon properties are determined model-independently
and are known to high precision phenomenologically
\cite{Li:1971vr}. For example, the LNA coefficient to the proton
magnetic moment is given by
\begin{equation}
\chi_{p\pi}=-\frac{g_A^2\,M_N}{8\,\pi\,\fpi^2} \, .
\end{equation}
Any extrapolation of lattice QCD simulations must incorporate this
knowledge of QCD.

The analytic terms are unconstrained by chiral symmetry, and hence
must be determined empirically. Lattice QCD provides an {\it ab
initio} framework to determine these parameters from QCD.

The expression, \eq{eq:exp}, is analogous to the chiral expansion of
the nucleon mass described in
Refs.~\cite{Young:2002ib,Leinweber:2003dg}.  The regulator dependence
of the integrals is implicit.  To define the expansion one must define
a regularisation and renormalisation scheme to remove ultra-violet
divergences.

In the text-book approach to elementary field theory one commonly uses
dimensional regularisation to renormalise the theory. In this case,
only the residue of the pion pole becomes apparent and the loop
contribution is therefore given by
\begin{equation}
\chi_{B} I_\pi^{\rm DR}               \rightarrow \chi_B \mpi\, ,
\end{equation}
and the linear divergence of the integral is absorbed into an 
infinite renormalisation of $a_0$.

Alternatively, ultra-violet divergences can be removed by suppressing
loop integrals above a finite energy scale. The first systematic study
of finite-range regularisation in \chpt\ was performed by Donoghue et
al.~\cite{Donoghue:1998bs}. The development of FRR-\chpt\ in the
context of lattice QCD
\cite{Young:2002ib,Leinweber:2003dg,Leinweber:1999ig,Leinweber:2001ac,Young:2002cj,Cloet:2003jm}
has found remarkably improved convergence properties of the chiral
expansion, providing access to a much wider
range of range of quark mass than the naively regularised theory
\cite{Young:2002ib}.  FRR is therefore best suited to the problem of
chiral extrapolation, where the systematic error in the extrapolation
of modern lattice simulations is of order 1\% \cite{Leinweber:2003dg}.

Using FRR, the loop integral, $I_\pi$, in the heavy baryon
limit can be expressed as
\begin{equation}
I_\pi = -\frac{4}{3\,\pi}\int dk \frac{k^4 u^2(k)}{\wk^4}\, ,
\label{eq:Ipi}
\end{equation}
where $\wk=\sqrt{k^2+\mpi^2}$ and $u(k)$ is the ultra-violet
regulator.

Expansion of the loop contributions as a power series enables one to
obtain the renormalised chiral expansion parameters, where each of the
analytic terms in \eq{eq:exp} are renormalised by a finite
amount \cite{Donoghue:1998bs,Young:2002ib}. For example, using a
dipole regulator, $u(k)=(1+k^2/\Lambda^2)^{-2}$, \eq{eq:Ipi}
becomes 
\begin{equation}
I_\pi^{\rm DIP} = - \frac{\Lambda^5 (\mpi + 5 \Lambda) }{24 (\mpi +
  \Lambda)^5 }\, ,
\label{eq:IpiDIP}
\end{equation}
and the Taylor expansion provides
\begin{equation}
I_\pi^{\rm DIP} = -\frac{5}{24}\Lambda + \mpi - \frac{35}{12\,\Lambda} \mpi^2 + \ldots
\end{equation}
Therefore, precisely the same LNA contribution is recovered
\begin{equation}
\chi_{B\pi} I_\pi^{\rm DIP\ {\rm (LNA)}} = \chi_{B\pi} \mpi\, ,
\end{equation}
with a finite renormalisation of all other terms in the series. By
varying $\Lambda$, strength in the loop integral may be moved to the
residual analytic expansion and vise-versa. As the moderately large
$\mpi$ behaviour of the loop integral and the residual expansion are
radically different, varying $\Lambda$ provides an opportunity to
optimize the convergence properties of the truncated chiral expansion.

We show the mathematical equivalence of the renormalisation
prescriptions to a given order.  The renormalised expansion in
dimensional regularisation is
\begin{equation}
\mu_B = c_0^B + \chi_{B\pi} \mpi + c_2^B \mpi^2 + \ldots \, ,
\end{equation}
and these renormalised coefficients  are recovered from the FRR
expansion via
\begin{eqnarray}
c_0^B &=& a_0^B - \chi_{B\pi}\frac{5}{24}\Lambda    \, , \non\\
c_2^B &=& a_2^B - \chi_{B\pi}\frac{35}{12\,\Lambda} \, ,
\end{eqnarray}
where the second terms compensate the $\Lambda$ dependence of $a_i^B$.

In summary, in working to leading order in the chiral expansion with a
dipole FRR, the quark mass dependence of nucleon magnetic moments in 
QCD is
\begin{equation}
\mu_B = a_0^B + a_2^B \mpi^2 + a_4^B \mpi^4 - \chi_{B\pi} \frac{\Lambda^5 (\mpi + 5 \Lambda) }{24 (\mpi +
  \Lambda)^5 }\, .
\label{eq:expDIP}
\end{equation}
The inclusion of an $\mpi^4$ term here is in anticipation of adding
NLNA terms in the following.


\section{Quenched Considerations}
Here we address the necessary modifications to the chiral effective
field theory for the quenched approximation. Vacuum fluctuations of
$q\bar{q}$-pairs are absent in quenched simulations. As a result, the
structure of the low-energy effective field theory is modified.
Meson loop diagrams are restricted to only
those where the loop is comprised of valence quarks having their
origin in the interpolating fields of the baryon correlation function.
This has the effect of modifying the
effective $\pi$--$N$ coupling constants
\cite{Savage:2001dy,Leinweber:2002qb} and the
corresponding factors $\chi_B$ are changed accordingly.

To summarize the LNA contributions to nucleon magnetic moments in both
quenched and dynamical QCD, we use the standard notation and define
\begin{equation}
\chi_{B\pi} = \frac{M_N}{8\pi\fpi^2}\beta_B^\pi \, ,
\label{eq:chiBpi}
\end{equation}
and provide the coefficients, $\beta_B^\pi$, are in
Table~\ref{tab:beta} \cite{Savage:2001dy,Leinweber:2002qb}. We use
$\fpi=93\mev$, $D=0.76$ and $F=0.50$, ($g_A=D+F$).

\begin{table}
\begin{center}
\caption{Coefficients of the leading pion-loop contributions to nucleon magnetic
  moments in QCD and QQCD.\label{tab:beta}}
\begin{ruledtabular}
\begin{tabular}{ccc}
Baryon    & $\beta_B^\pi$  & $\beta_B^{\pi\, {\rm (Q)}}$ \\
\noalign{\smallskip}
\hline
$p$       & $-(F+D)^2$     & $-\frac{4}{3}D^2$           \\
$n$       & $(F+D)^2$      & $\frac{4}{3}D^2$            \\
\end{tabular}
\end{ruledtabular}
\end{center}
\end{table}

A peculiar feature of the quenched theory is the appearance of the
flavour-singlet \npr-meson as a light degree of freedom. In the
absence of vacuum quark loops the \npr\ behaves as a Goldstone boson
\cite{Sharpe:1992ft,Bernard:1992mk} and must therefore be incorporated
in the low-energy effective field theory.

\begin{figure}[!b]
\begin{center}
\includegraphics[width=5cm, angle=0]{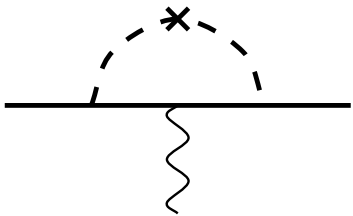}
\caption{Double-hairpin \npr\ vertex correction to nucleon magnetic moment.
\label{fig:eta}}
\end{center}
\end{figure}

The \npr\ carries no charge and therefore does not make a direct
contribution to the magnetic moment of the nucleon. The leading
enhancement of the moment comes from \npr-dressing of the current
insertion as in Fig.~\ref{fig:eta}.
Being a flavour singlet, the \npr\ can propagate through a pure
gluonic state.  This corresponds to a double-hairpin in
the quark-flow diagram.
The vertex correction to the magnetic moment induced by this loop
produces more singular nonanalytic behaviour in the chiral limit than
the physical theory.

The double-hairpin diagram has a logarithmic divergence near the
chiral limit. This is a pathological feature of the quenched
approximation, where the magnetic moment tends to infinity near the
chiral limit. Physically this is not possible as angular momentum
quantization ensures that the induced magnetic field of the meson-loop
remains finite, even in the chiral limit
\cite{Leinweber:2001ui}.

The double-hairpin vertex correction diagram has been calculated in the
graded-symmetry approach to Q\chpt\ by Savage in
Ref.~\cite{Savage:2001dy}, and provides the term
\begin{equation}
\chi_{\eta '}^{\rm (Q)} \mu_B^{\rm (Q) tree} I_{\eta '} \, ,
\end{equation}
with coefficient
\begin{equation}
\chi_{\eta '}^{\rm (Q)} = \frac{m_0^2\, (3 F - D)^2}{12\,\pi^2\,\fpi^2}\, .
\end{equation}
The superscript (Q) denotes a quenched quantity.  The coefficient
$m_0^2$ is associated with the double-hairpin vertex in the \npr\
propagator. This parameter is related to the physical \npr\ mass
\cite{Sharpe:1992ft,Bernard:1992mk} and we choose a value
$m_0^2=0.42\gev$.  The loop integral in the heavy-baryon limit is
given by
\begin{equation}
I_{\eta '} = -\int dk \frac{k^4 u^2(k)}{\wk^5}\, .
\label{eq:Ietapr}
\end{equation}
The normalization of the integral is such that the LNA contribution to
this loop is $\log \mpi$. 

We also highlight that this contribution is proportional to the
tree-level moment, $\mu_B^{\rm (Q) tree}$ \cite{Savage:2001dy}. Because of the logarithmic
divergence one cannot simply define this to be the renormalised moment
in the chiral limit. This necessarily means that one cannot remove the
scale dependence of the coefficient of this chiral log using standard
methods. We remove the scale dependence by replacing the tree level
coefficient by the renormalised magnetic moment at each quark mass
$\mu_B^{\rm (Q) tree}=\mu_B^{\rm (Q)}$. This approximation will be
accurate provided $\chi_{\eta '}^{\rm (Q)} \mu_B^{\rm (Q) tree}
I_{\eta '}$ makes only small contributions for $\mpi>\mpi^{\rm
  phys}$.

\begin{figure}[!t]
\begin{center}
\includegraphics[width=\hsize]{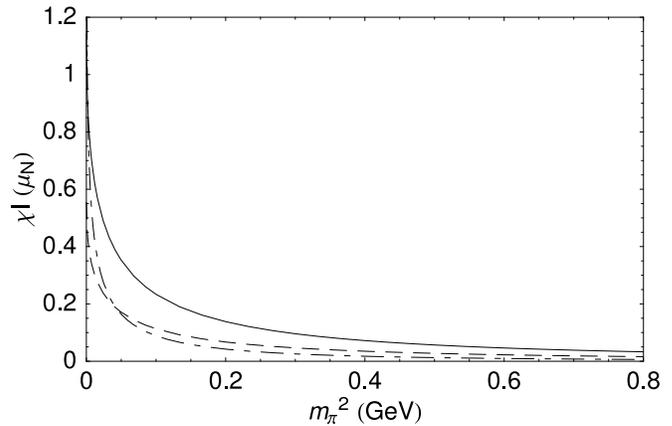}
\end{center}
\caption{Loop corrections, in units of $\NM$, evaluated with a dipole
  FRR with $\Lambda=0.8\ \gev$. The dashed and solid curves are the
  pion loop corrections of Fig.~\ref{fig:piLNA} in quenched and
  physical QCD respectively. The dash-dot curve is the \npr\ vertex
  correction of quenched QCD, where the physical magnetic
  moment is used to define the renormalized coupling.
\label{fig:loopval}}
\end{figure}

In Figure~\ref{fig:loopval} we show the value of
the loop contributions for varying pion mass.  The corrections from
the \npr\ are quite small in the region of interest.

Our expansion in the quenched approximation, analogous to \eq{eq:exp},
is given by
\begin{eqnarray}
\mu_B^{\rm (Q)} &=& a_0^{B\, \rm (Q)} + a_2^{B\, \rm (Q)} \mpi^2 +
      a_4^{B\, \rm (Q)} \mpi^4 \non\\
      & & + \chi_{B\pi}^{\rm (Q)}\, I_\pi 
          + \chi_{B\eta '}^{\rm (Q)}\, \mu_B^{\rm (Q)}\, I_{\eta '} \, ,
\label{eq:qexp}
\end{eqnarray}
and is used to determine the parameters $a_i^{B\, \rm (Q)}$.  The total
magnetic moment at arbitrary $\mpi$ is then given by
\begin{eqnarray}
\mu_B^{\rm (Q)} &=& 
                \left\{ a_0^{B\, \rm (Q)} + a_2^{B\, \rm (Q)} \mpi^2
                + a_4^{B\, \rm (Q)} \mpi^4 \right . \non\\
                &&\quad + \left . \chi_{B\pi}^{\rm (Q)} I_\pi \right\} \, 
\left(1 - \chi_{B\eta '}^{\rm (Q)} I_{\eta '}\right)^{-1} \, .
\label{eq:qfit}
\end{eqnarray}
We note that at this point the decuplet contributions have been
suppressed as their contributions are higher-order in the chiral
expansion, when one accounts for the octet-decuplet mass splitting
realized in the (quenched) chiral limit \cite{Young:2002cj}.


\section{Extrapolation of Lattice Magnetic Moments}

\begin{figure}[!t]
\begin{center}
\includegraphics[width=\hsize]{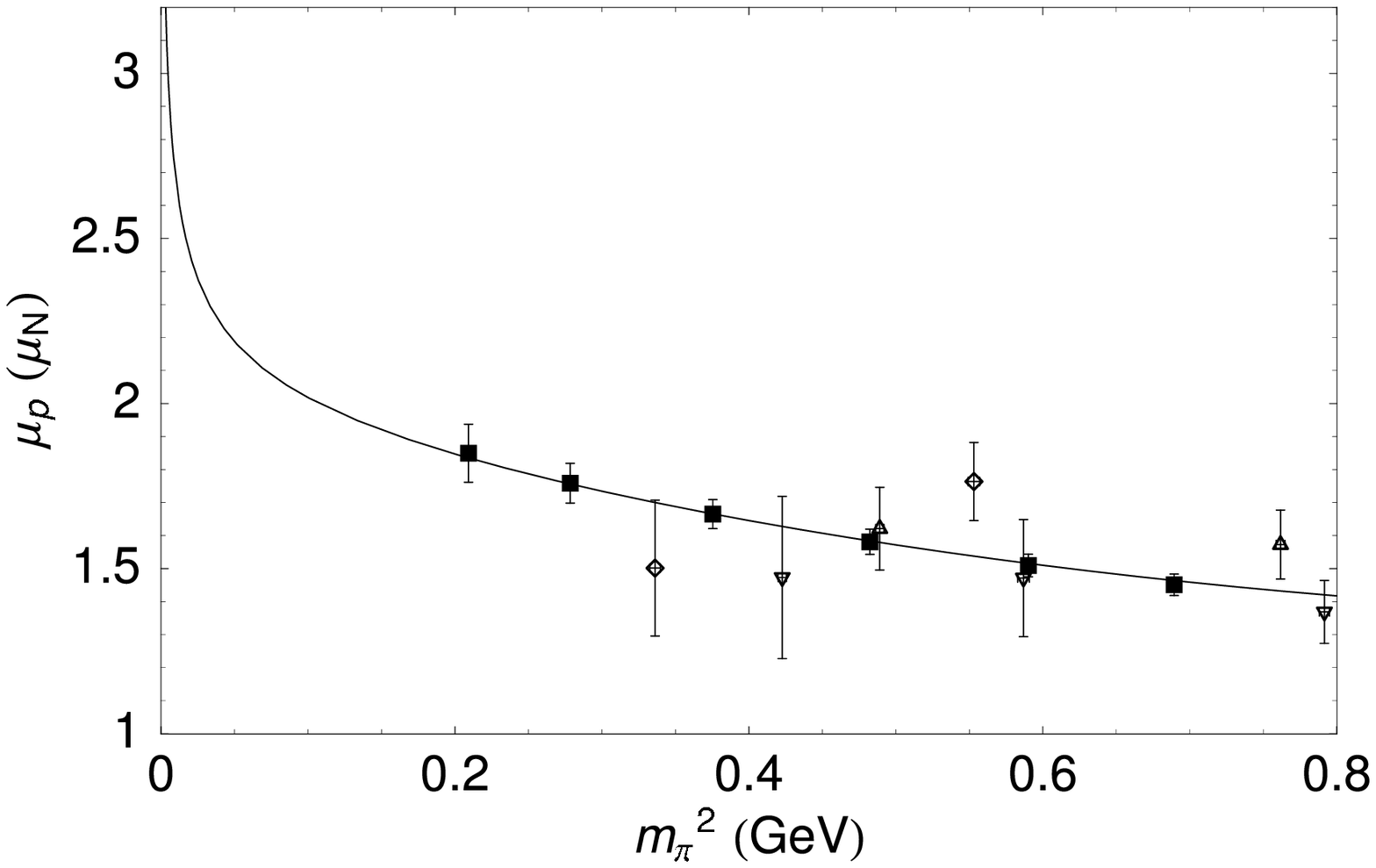}
\end{center}
\caption{Fit of \eq{eq:qexp} to quenched lattice data. The solid
squares ($\blacksquare$) illustrate the FLIC fermion results
\cite{Zanotti:2003gc,Zanotti:2004jy} and the open symbols describe
nonperturbatively improved clover results \cite{Gockeler:2003ay}, at
$\beta=6.0$ ($\triangledown$), $\beta=6.2$ ($\triangle$) and
$\beta=6.4$ ($\lozenge$).}
\label{fig:qfit}
\end{figure}

With the expansion of the low-energy EFT determined, one needs to fix
the values of the unconstrained terms empirically. Lattice QCD
provides information on the quark mass dependence of hadronic
properties and hence enables the determination of these free
parameters. If the data analyzed is within the applicable range of the
effective field theory then, upon fixing these low-energy constants,
one has an accurate extrapolation to the physical regime.

The electromagnetic form factors of the nucleon have recently been
studied in simulations of quenched lattice QCD
\cite{Gockeler:2003ay,Zanotti:2003gc,Zanotti:2004jy}.  Early
simulations of nucleon three-point functions have been performed by
Leinweber {\it et al.} in Ref.~\cite{Leinweber:1991dv}.  Direct
lattice calculations of the nucleon's strangeness form factor have
also been carried out in Refs.~\cite{Dong:1997xr,Lewis:2002ix}.

In this study we choose to analyze only the most recently performed
simulations using improved quark actions. Results by Gockeler et
al.~\cite{Gockeler:2003ay} have been obtained using the
nonperturbatively improved clover fermion action
\cite{Luscher:1996ug}.  We also consider recent form factor
simulations by Zanotti et al.~\cite{Zanotti:2003gc,Zanotti:2004jy}
using the fat-link irrelevant clover (FLIC) quark action
\cite{Zanotti:2001yb}. For the purposes of this investigation, we
select the six most accurate data points from the FLIC data set. The
precision of the FLIC fermion results reflects the use of improved
unbiased estimation techniques \cite{Leinweber:1991dv}, improved
actions and high statistics.  It has been demonstrated that
FRR-$\chi$EFT is applicable up to $\mpi^2=0.8\gev^2$
\cite{Young:2002ib} and hence we also choose to truncate our data set
at this scale of pion mass.

All previous chiral extrapolations of lattice electromagnetic
structure have been based on \chpt\ under the assumption that
quenching effects are minimal
\cite{Leinweber:1998ej,Leinweber:1999nf,Hackett-Jones:2000qk,%
Hackett-Jones:2000js,Hemmert:2002uh,Cloet:2003jm,Gockeler:2003ay,%
Ashley:2003sn,Flambaum:2004tm}.
Here we present the first comprehensive analysis of quenched lattice
magnetic moments using Q\chpt\ for baryon form factors.  Brief reports
on preliminary Q\chpt\ extrapolations have recently been presented in
Refs.~\cite{Leinweber:2003ux,Young:2003gd,Young:2003ns}.

In Fig.~\ref{fig:qfit} we show fits to quenched lattice data of the
proton magnetic moment using \eqs{eq:qexp}{eq:qfit}.  In anticipation
of estimating unquenching corrections we have adopted the preferred
value of $\Lambda=0.8\gev$ \cite{Young:2002cj}. The logarithmic
divergence of $\mu_p$ in the chiral limit is evident.

\begin{figure}[!t]
\begin{center}
\includegraphics[width=\hsize]{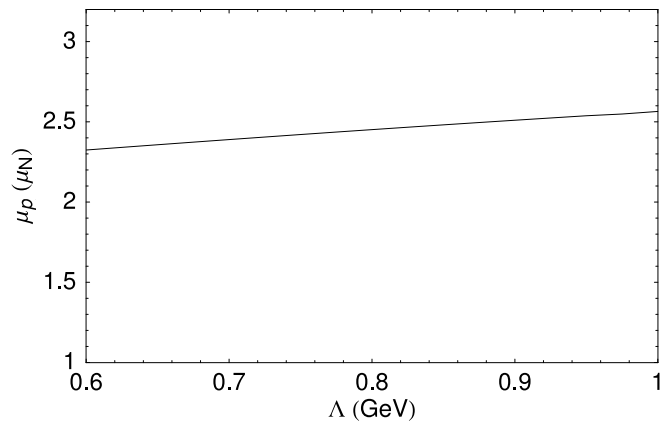}
\end{center}
\caption{The dependence of the extrapolated magnetic moment determined
  at the physical pion mass on the dipole regulator parameter
  $\Lambda$.  Variation in the quenched proton moment is similar to
  the statistical uncertainty of the FLIC simulation result at the
  lightest quark mass considered.
\label{fig:lam}}
\end{figure}

Fig.~\ref{fig:lam} illustrates the dependence of the extrapolated
result, evaluated at the physical pion mass, on the choice of
regulator parameter $\Lambda$. The vertical axis has been fixed to
that of Fig.~\ref{fig:qfit} in order to display the relevant scale.
The variation in the quenched proton moment is similar to the
statistical uncertainty of the FLIC simulation result at the lightest
quark mass considered.  Further reduction of the $\Lambda$ sensitivity
could be obtained by considering higher-order terms of the chiral
expansion. 

\section{Extrapolation on a Finite Volume}
The previous section assumed the lattice QCD results provide an
accurate representation of the results to be realized in the
infinite-volume limit.  However, results obtained on a lattice of
finite spatial extent will differ from those in the infinite volume
limit, particularly in the chiral regime where the pion Compton
wavelength can approach the lattice length.  Here we extend the
formalism of FRR to incorporate finite-volume effects.  

The leading-order finite-volume effect in the chiral expansion lies in
terms analytic in the small expansion parameter, $1/L$, where $L$ is
the length of the cubic volume~\cite{Gasser:1988zq}.  Finite-volume
corrections also enter through the modification of loop integrals.
The requirement that Greens functions are periodic
\cite{Gasser:1988zq} restricts momentum components to the values
\begin{equation}
k_i = \frac{2 \pi n_i}{L}\, , \quad  n_i = 0,\ \pm 1,\ \pm 2,\ 
\ldots \, . 
\end{equation}
For the $p$-wave loop contribution of Fig.~\ref{fig:piLNA}, where
strength in the integrand vanishes for $k = 0$ as in
Eq.~(\ref{eq:Ipi}), the dominant effect of discretizing the momenta is
to introduce a threshold effect \cite{Leinweber:2001ac,Young:2002cj}.
Strength in the integrand is not sampled until one component of $k$
reaches $2 \pi / L$.  Since chiral physics is dominated by the
infrared behaviour of loop integrals, the nonanalytic terms of the
chiral expansion exhibit substantial threshold effects.

To obtain a complete description of the quark mass and volume
dependence of hadron properties, one must have an expansion in both
$m_q$ and $1/L$ \cite{Gasser:1988zq}.  Without data at varying lattice
volumes it is impossible to determine the expansion coefficients of
the $1/L$ contributions. Recent calculations have also considered the
finite volume corrections arising from the modification of the leading
loop integrals to discretised momentum sums
\cite{AliKhan:2003cu,Beane:2004tw}.  Impressive results for a
phenomenological description of the volume dependence of hadron masses
have also recently been reported \cite{Orth:2003nb}.

In the absence of lattice QCD results for magnetic moments from a
variety of lattice volumes, it is not possible to rigorously constrain
the complete volume-dependent expansion.  However, it is possible to
precisely describe the impact of the finite volume on the quark-mass
dependence of magnetic moments on a single, fixed lattice volume.

Evaluating the loop integrals on a finite volume in FRR is a rather
simple extension.  One simply replaces the continuum integral by the
discrete momentum sum over available momenta on a given volume.  We
formulate the finite-volume corrections in the continuum theory and
therefore continue the momentum sum to infinity.  This allows the
features of finite-range regularization to carry over to the finite
volume case.  For example, \eq{eq:Ipi} becomes
\begin{eqnarray}
I_\pi &=& -\frac{4}{3\,\pi}\int dk\, \frac{k^4\, u^2(k)}{\wk^4}
      = -\frac{1}{3\,\pi^2}\int d^3 k\, \frac{k^2\, u^2(k)}{\wk^4} 
\nonumber \\
      &\rightarrow& -\frac{1}{3\,\pi^2} \left (\frac{2\pi}{L} \right )^3
      \sum_{\vec{k}} \frac{k^2\, u^2(k)}{\wk^4} \, .
\label{eq:IpiDisc}
\end{eqnarray}
The discretized momenta on a cube are given by $\vec{k}=k_{\rm
min}\vec{\mathbf n}$ for $\vec{\mathbf n}\in \mathbb{Z}^3$, with the
minimum nontrivial momentum given by $k_{\rm min}=2\pi/L$. We note
that in obtaining the infinite volume form of Eq.~(\ref{eq:Ipi}) the
angular dependence of the 3-dimensional integral has been performed
analytically. In general, with an angular dependent integrand, one
must take caution in the naive conversion to a spherically symmetric
sum as shown in Eq.~(\ref{eq:IpiDisc}). For the integrals used in this
paper, we have verified that the spherically symmetric and angular
dependent summations are equivalent.

Simulations performed with standard Wilson actions have large ${\cal
O}(a)$ errors that can be accounted for in the effective field theory.
One can introduce new local operators into the chiral Lagrangian
reflecting ${\cal O}(a)$ terms associated with lattice discretization
effects and explicit chiral symmetry breaking 
\cite{Rupak:2002sm,Bar:2002nr,Aoki:2003yv}.
However, there has been tremendous success in removing ${\cal O}(a)$
errors and suppressing ${\cal O}(a^2)$ errors in lattice simulation
results through the development of nonperturbatively improved actions
\cite{Luscher:1996sc,Zanotti:2001yb,overlap}.  These actions display
excellent scaling properties
\cite{FLICscaling,Edwards:1998nh,Dong:2000mr}, providing near
continuum results at finite lattice spacing.  In particular, the FLIC
fermion simulations, dominating the chiral fits here, are performed
using an ${\cal O}(a^2)$--mean-field improved Luscher-Weisz plaquette
plus rectangle gauge action \cite{Luscher:1984xn} and the
nonperturbatively ${\cal O}(a)$-improved FLIC fermion action
\cite{Zanotti:2001yb,FLICscaling}.  Hence these lattice results
already represent an excellent approximation to the continuum limit.

\begin{figure}[!t]
\begin{center}
\includegraphics[width=\hsize]{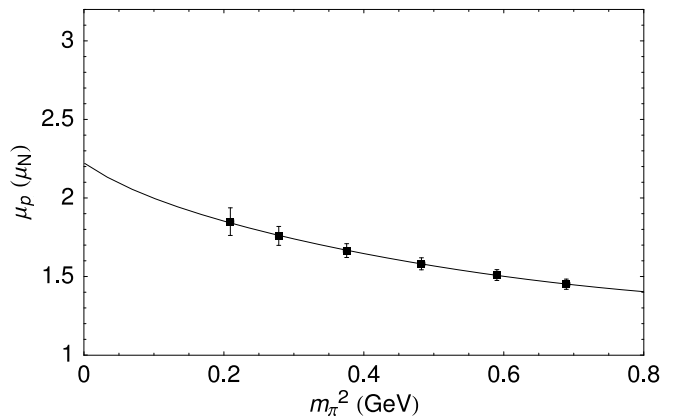}
\end{center}
\caption{Finite-volume FRR-EFT fit to FLIC fermion results for fixed
volume.  Here the dipole-vertex regulator parameter is fixed to
$\Lambda=0.8\gev$.}
\label{fig:qfitvol}
\end{figure}

\begin{figure}[!t]
\begin{center}
\includegraphics[width=\hsize]{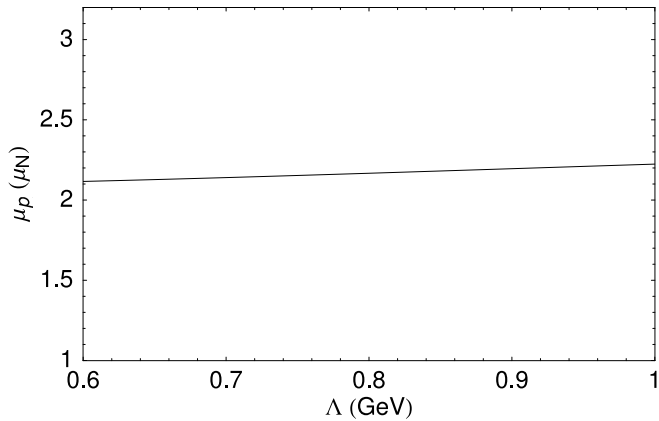}
\end{center}
\caption{The extrapolated magnetic moment on a finite volume,
$V=(2.56\fm)^3$, for varying regulator parameter $\Lambda$.
\label{fig:lamvol}}
\end{figure}

Fig.~\ref{fig:qfitvol} illustrates the fit of Eq.~(\ref{eq:qexp}) with
the $\pi$ and $\eta'$ loop integrals modified as described in
Eq.~(\ref{eq:IpiDisc}) for the finite volume.  The physical volume of the FLIC
lattice is $V=(2.56\fm)^3$.  The chiral properties of the
finite-volume extrapolation are qualitatively different from the
infinite volume curve of Fig.~\ref{fig:qfit}.  In
Fig.~\ref{fig:lamvol} we show the regulator parameter dependence on
the finite-volume extrapolated magnetic moment for $\Lambda$ in the
range $0.6$--$1.0\gev$.  The variation of the extrapolated moment is
suppressed relative to the infinite volume case, changing by $0.1\
\mu_N$.  This systematic uncertainty is smaller than the statistical
uncertainty of the lightest quark mass considered here, and could be
suppressed further through the introduction of higher-order terms in
the chiral expansion or through the introduction of precise lattice
QCD results at light quark masses.

\section{Estimating Effects of Dynamical Sea Quarks}

In a study of baryon masses in quenched and 2+1-flavour QCD it has
been found that the short-distance physics of the analytic terms in
the residual expansion of FRR-EFT are found to be very similar when
the chiral-loop effects are evaluated with an appropriate FRR
\cite{Young:2002cj}.  By identifying the short distance behaviour in
QQCD, one need only restore the chiral loop effects of QCD to obtain
an improved estimate of the physical magnetic moment.

In making such an identification it is essential to have a consistent
method for setting the scale in both quenched and dynamical QCD. In
particular, one must ensure that the procedure is insensitive to
chiral physics. The QCD Sommer scale \cite{Edwards:1997xf},
based on the static quark potential, is insensitive to light quark
physics and provides an ideal procedure for the scale
determination.

The identification of this phenomenological link between quenched and
dynamical simulations has been applied to FLIC fermion calculations of
baryon masses \cite{Leinweber:2003ux,Young:2003ns}. Upon replacing the
chiral loops of QQCD by their QCD counterparts the nucleon and Delta
are found to be in good agreement with experiment.

\begin{figure}[!t]
\begin{center}
\includegraphics[width=\hsize]{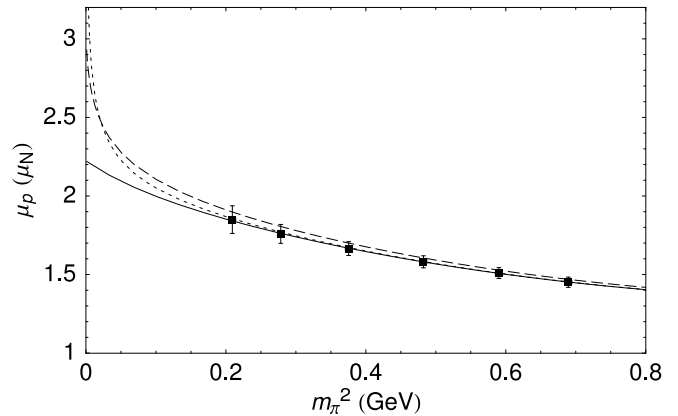}
\end{center}
\caption{Correcting the finite-volume quenched approximation to the
infinite-volume limit of full QCD.
The solid curve is the finite-volume quenched fit to the data as in
Fig.~\ref{fig:qfit}.
The dotted curve provides an estimate of the infinite-volume limit
magnetic moment in the quenched approximation.
The dashed curve shows estimates of the proton magnetic moment in full
QCD as described in the text.
\label{fig:magQCD}}
\end{figure}

By applying the same principle to the calculation of magnetic moments
in quenched QCD one can obtain improved estimates of the physical
magnetic moment.  The fit parameters, $a_i^{B\, \rm (Q)}$, are
determined by fitting finite-volume quenched lattice QCD using
\eq{eq:qexp} with discretized momenta and a dipole regulator of
$0.8\gev$.  The estimate of the quenching effects are obtained under
the assumption that the bare residual expansion parameters are
unchanged in infinite-volume QCD when $\Lambda = 0.8 \gev$.  That is,
the full QCD result can be described by \eq{eq:expDIP} with the
identification $a_i^{B\, \rm (Q)}=a_i^{B}$.  By fitting with
finite-volume FRR-EFT both quenching and finite-volume corrections are
incorporated in the final estimate.  We show the infinite-volume QCD
estimate of the proton magnetic moment by the dashed curve in
Fig.~\ref{fig:magQCD}.

In a similar manner, the infinite-volume limit of QQCD is estimated by
fitting the parameters $a_i^{B\, \rm (Q)}$ of \eq{eq:qexp} using
finite-volume discretized momenta and a dipole regulator of $0.8\gev$
in the loop integrals.  The correction is estimated by \eq{eq:qfit}
calculated with infinite-volume continuous momenta in the loop
integrals.  Figure~\ref{fig:magQCD} illustrates that the finite volume
corrections are negligible in the regime of the lattice QCD simulation
results. 

We emphasize that this result is a phenomenological estimate, as the
size of the correction is $\Lambda$ dependent.  However an important
feature of this approach is that the largest finite volume corrections
lie in the chiral limit as they should.  Ultimately, one would like to
combine the improved convergence properties of FRR-EFT with the small
$1/L$ expansion such that accurate and model-independent
determinations of finite volume effects can be made.

The primary feature of Fig.~\ref{fig:magQCD} is that although the
quenched and physical theory have quite different chiral structure,
the observable effects are rather small.  In particular, the
logarithmic divergence is likely to only become apparent well below
the physical pion mass.  Within the current formalism of lattice QCD it
seems such observation would be a formidable task, particularly given
the large lattice volume required to reveal the \npr contribution.

The results here, based on the leading chiral corrections, indicate
that proton magnetic moments evaluated in quenched simulations give a
good approximation to the true theory.  The enhancement from
the \npr-loop compensates for the reduction in the standard pion-loop
from QCD to QQCD.  The similarity in the effective curvature was also
highlighted by Savage \cite{Savage:2001dy}.


\section{Series Truncation and Higher Order Effects}

It is interesting to explore the convergence properties of the series
expansion, generated in dimensional regularisation, truncated to
various order in $\mpi$.  Information on the convergence of the
truncated series can be obtained by studying the Taylor series
expansion of the FRR dipole regularised form, \eq{eq:expDIP}, which
describes lattice results very well. Determination of the convergence
of the series will therefore give insight into the applicable range of
a dimensionally regularised expansion. Similar studies of a truncation
of the chiral loop corrections to the nucleon mass evaluated with a
FRR have been performed in
Refs.~\cite{Stuckey:1997qr,SVW,Bernard:2002yk}.

\begin{figure}[!t]
\begin{center}
\includegraphics[width=\hsize]{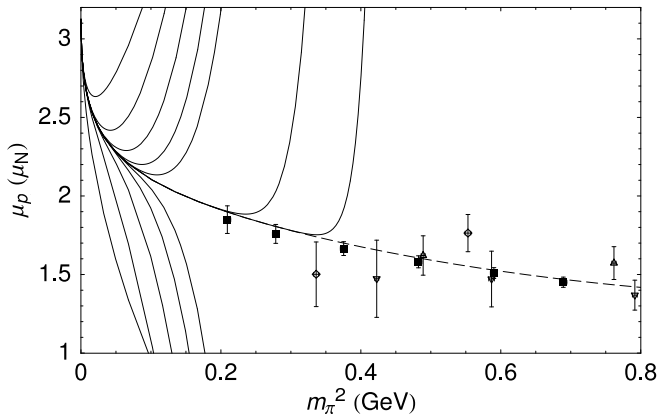}
\end{center}
\caption{Truncations of the Taylor expansion of the full QCD
expression of the magnetic moment, \eq{eq:expDIP}, to various orders
in $\mpi$. The left most curves display a succession of truncations
from $\mpi$ up to $\mpi^{10}$. To reach convergence at the lightest
FLIC data point considered here one requires terms up to
$\mpi^{26}$, while for QCDSF terms to $\mpi^{50}$ are necessary. The 
dashed curve displays the full QCD expression as displayed in 
Fig.~\ref{fig:magQCD}.
\label{fig:trunc}}
\end{figure}

In Fig.~\ref{fig:trunc} we show the Taylor series expansion of
\eq{eq:expDIP} truncated at various powers of $\mpi$. This plot
indicates that in order to accurately reproduce the quark mass
dependence of the proton magnetic moment up to $\mpi\sim0.1\gev^2$
then one must keep all terms up to $\mpi^{10}$, {\it i.e.} $m_q^5$. If
one wishes to reach the quark mass scale of modern lattice QCD
simulations many more powers in $\mpi$ are required.
In fact, the lightest displayed FLIC quark mass requires terms to
$\mpi^{26}$, similarly the lightest QCDSF point needs $\mpi^{50}$.

This problem might have been anticipated by the fact that at
moderately large quark masses the Dirac moment of the nucleon would be
revealed
\begin{equation}
\mu_p = \frac{e\hbar}{2 M_N}\, .
\end{equation}
Knowing that in this regime, the nucleon mass grows linearly with
$\mpi^2$, the moment would require an expansion in inverse powers of
$\mpi^2$ to describe the data
\cite{Leinweber:1998ej,Hackett-Jones:2000qk}. 

\begin{figure}[!t]
\begin{center}
\includegraphics[width=5cm, angle=0]{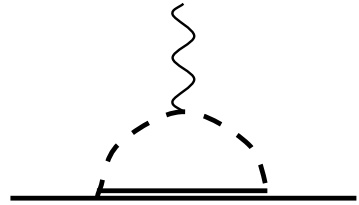}
\caption{Leading pion-loop contribution to nucleon magnetic moment
from the $\Delta$ resonance.
\label{fig:delta}}
\end{center}
\end{figure}
We also investigate some of the effects of including higher-order
terms in the FRR extrapolation of lattice data.  The leading contribution
from the Delta baryon, shown in Fig.~\ref{fig:delta}, is also
incorporated into the fit. Equation (\ref{eq:qexp}) becomes
\begin{eqnarray}
\mu_B^{\rm (Q)} &=& a_0^{B\, \rm (Q)} + a_2^{B\, \rm (Q)} \mpi^2
                 + a_4^{B\, \rm (Q)} \mpi^4 \non \\
             &&  + \chi_{B\eta '}^{\rm (Q)}\, \mu_B^{\rm (Q)}\, I_{\eta '} 
                 + \chi_{B\pi}^{\rm (Q)}\, I_\pi 
                 + \chi_{B\pi}^{\Delta\rm (Q)}\, I_\Delta\pi \, ,
\label{eq:qfitD}
\end{eqnarray}
where, analogous to \eq{eq:chiBpi}, the corresponding couplings are given by
\begin{equation}
\chi_{B\pi}^\Delta = \frac{M_N}{8\pi\fpi^2}\, \beta_{B\Delta}^\pi \, ,
\end{equation}
and Table \ref{tab:betaD}. The loop integral is also modified by the
fact that the intermediate baryon propagator is non-degenerate with
the external state.  With the mass-splitting given by $\Delta$,
\eq{eq:Ipi} becomes
\begin{equation}
I_\pi = -\frac{4}{3\,\pi}\int dk \frac{(\Delta+2\wk)k^4 u^2(k)}{2\wk^3(\Delta+\wk)^2}\, .
\end{equation}

\begin{table}
\begin{center}
\caption{Coefficients of the leading decuplet, pion-loop contributions to nucleon magnetic
  moments in QCD and QQCD \cite{Savage:2001dy}, $\C=-2 D$.
\label{tab:betaD}}
\begin{ruledtabular}
\begin{tabular}{ccc}
Baryon    & $\beta_{B\Delta}^\pi$  & $
\beta_{B\Delta}^{\pi\, {\rm (Q)}}$ \\
\noalign{\smallskip}
\hline
$p$       & $-\frac{2}{9}\C^2$     & $-\frac{1}{6}\C^2$           \\
$n$       & $\frac{2}{9}\C^2$      & $\frac{1}{6}\C^2$            \\
\end{tabular}
\end{ruledtabular}
\end{center}
\end{table}

Given that the finite-volume corrections are negligible in the regime
where the lattice QCD results lie, we illustrate the role of the
$\Delta$ in FRR-EFT by taking the lattice results as an accurate
representation of the infinite-volume limit and evaluate the
loop integrals of \eq{eq:qfitD} in the infinite volume limit.

\begin{figure}[!t]
\begin{center}
\includegraphics[width=\hsize]{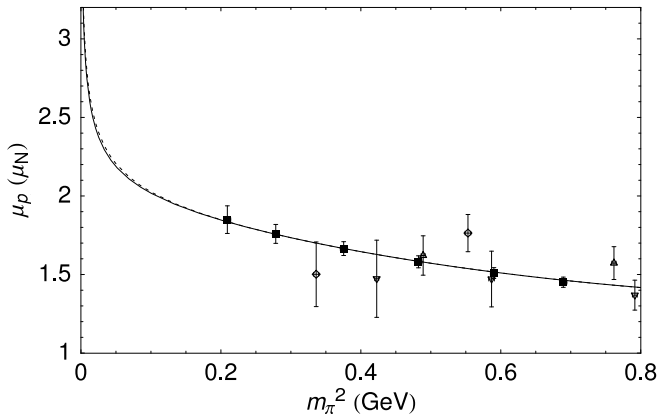}
\end{center}
\caption{The role of $\Delta$ contributions in the extrapolation of
  the proton magnetic moment.  The solid curve displays the
  leading-order QQCD extrapolation of \eq{eq:qfit}, without explicit
  decuplet contributions, as seen in Fig.~\ref{fig:qfit}.  If
  \eq{eq:qfit} is extended to include the pion contributions with
  decuplet baryons in \eq{eq:qfitD}, one obtains the dotted curve.
\label{fig:delt}}
\end{figure}

The fit of \eq{eq:qfitD} to lattice results is shown by the dotted
curve of Fig.~\ref{fig:delt}.  The comparison with the leading order
result (solid curve) shows that the effect of the decuplet on the
extrapolation result is negligible.  The new non-analytic behaviour
introduced by an explicit inclusion of the $\Delta$ was already
approximated well by the analytic terms in the expansion at leading
order.  Although the extrapolation shows little sensitivity to the
inclusion of the $\Delta$, it will be necessary to explicitly include
this degree of freedom if one is to extract the low-energy constants
to this order.

With regard to the estimation of full QCD corrections, the inclusion
of the decuplet is again a small effect.  With the $\Delta$, the QCD
estimate of the proton magnetic moment at the physical quark mass is
increased by $0.05\ \NM$ from the leading-order result.  Although small
for the proton, the inclusion of such contributions is found to be
important for other baryons of the octet in the extraction of the
strangeness magnetic moment of the proton
\cite{Leinweber:2003zy,strange}.

Finally, the prediction of the physical proton magnetic moment
obtained by including the $\Delta$ contributions and compensating for
the finite lattice volume is $2.54(30)\ \NM$, where the uncertainty is
statistical in origin.  This result agrees well with the experimental
value of $2.79\ \NM$.


\section{Conclusions}
Quenched and physical magnetic moments are in good agreement over a
large range of pion mass. Although pion effects alone are decreased in
QQCD, the new behaviour introduced by the \npr\ acts to suppress any
difference.

It is commonly accepted that the dimensionally regularised expansion
of the nucleon mass to ${\cal O}(p^4)$ is convergent to
$\mpi\sim 300\mev$. We find that to reach similar pion masses in the
DR expansion of the magnetic moment one would require knowledge of the
chiral expansion to ${\cal O}(p^{10})$.  The smooth behaviour of the
lattice data, together with the series truncations of the FRR
expansion indicate that although higher-order terms of DR can be
individually large they effectively sum to zero in the region of
interest.  FRR-EFT provides an effective resummation of the chiral
expansion that ensures that the slow variation of magnetic moments
observed in lattice QCD arises naturally in the FRR expansion.

Finally, by estimating finite volume and quenching effects through the
leading one-loop contributions of the finite-range regularized meson
cloud, we obtain an excellent value for the physical magnetic moment
of the proton.  Combined with the previous success in describing the
difference between quenched and full-QCD nucleon and $\Delta$ masses
\cite{Young:2002cj}, these results strengthen the argument that
artifacts of the quenched approximation can be accurately corrected
using phenomenological methods.  Ultimately, lattice QCD simulations
incorporating light dynamical-quark degrees of freedom on a variety of
lattice volumes are required to achieve the goal of an {\it ab initio}
determination of nucleon properties.


\acknowledgments

We would like to thank Ian Cloet, Emily Hackett-Jones and Stewart
Wright for helpful discussions.  This work was supported by the
Australian Research Council and by DOE contract DE-AC05-84ER40150,
under which SURA operates Jefferson Laboratory.


\end{document}